\documentstyle[12pt]{article}
\newcommand{\f}{\begin{equation}}
\newcommand{\ff}{\end{equation}}
\newcommand{\p}{{}^{\scriptstyle{+}}\!}
\newcommand{\m}{{}^{\scriptstyle{-}}\!}
\newcommand{\ppmm}{{}^{\scriptstyle{\pm}}\!}
\newcommand{\x}{  (\vec x)  }
\newcommand{\k}{  \vec k  }

\begin{document}

\vfil
\centerline{\bf A Loop Representation for the Quantum Maxwell Field}
\vfill
\centerline{Abhay Ashtekar${}^{1,2}$ and Carlo Rovelli${}^{3,4}$}
\medskip
\centerline{${}^1$ {\it Physics Department, Syracuse University, Syracuse,
NY 13244-1130, USA.}}
\centerline{${}^2$ {\it Theoretical Physics, Imperial College, London
SW7 2BZ, UK.}}
\centerline{${}^3$ {\it Physics Department, University of Pittsburgh,
Pittsburgh, PA 15260, USA.}}
\centerline{${}^4${\it Dipartimento di fisica, Universit\'a di Trento,
38050 Povo, Trento, Italia.}}
\vfill

Quantization of the free Maxwell field in Minkowski space is carried out using
a loop representation and shown to be equivalent to the standard Fock
quantization. Because it is based on coherent state methods, this framework may
be useful in quantum optics. It is also well-suited for the discussion of
issues related to flux quantization in condensed matter physics. Our own
motivation, however, came from a non-perturbative approach to quantum gravity.
The concrete results obtained in this paper for the Maxwell field provide
independent support for that approach. In addition, they offer some insight
into the physical interpretation of the mathematical structures that play,
within this approach, an essential role in the description of the quantum
geometry at Planck scale.

\nonumber\vfill\break

\section{Introduction.}

This paper discusses only the free quantum Maxwell field.
However, the motivation for the work comes from certain issues
that arise in quantum gravity.

Over the last decade, it has become increasingly clear that
perturbative methods, which have been so successful in the treatment
of non-gravitational interactions, cannot be used to {\it construct} a
quantum theory of gravity. One must face this problem
non-perturbatively. The canonical approach is well-suited for this
task. Indeed, already in the sixties, through the work initiated by
Bergmann, Dirac, Arnowitt, Deser, Misner and others, a concrete,
non-perturbative quantization program was formulated within this
approach. The idea was to first represent quantum states as functionals
of $3$-metrics and then select physical states by demanding that they
be annihilated by the quantum constraint operators. Unfortunately,
the task of solving the constraint equations turned out to be difficult
in the quantum theory and not a single physical state could be obtained.
More recently, it was realized that the problem simplifies considerably
if, prior to quantization, one performs a transform and works with
new canonical variables which are analogous to those normally used in
the Yang-Mills theory \cite{1}. Thus, in this framework, the
emphasis is shifted from $3$-metrics to certain connections on
the spatial $3$-manifold. Consequently, concepts such as the trace
of holonomies --the Wilson loops-- now play an important role in
quantum gravity. In fact, one can introduce \cite{2} a new representation
for quantum states  --called {\it loop representation}--  where states
are functionals of closed loops. This representation has turned out to
be particularly useful in solving the quantum constraints; an infinite
dimensional space of solutions is now known \cite{2,3}. We thus have a
large class of physical states of exact quantum gravity which can be used
to analyze the micro-structure of space-time in the Planck regime. What
we lack, however, is sufficient physical intuition for these states and
for the natural mathematical operators which operate on them.
There does exist a formal transform, called the {\it loop transform},
which enables one to relate objects on the loop space to those on the
space of connections which are generally easy to interpret geometrically
and hence also physically. However, the transform is only formal: it
involves an integration on the infinite dimensional space of connections
and very little is known about the existence of the required measure.
\footnote{The Gelfand spectral theory offers a promising approach to
making the transform well-defined \cite{4}.}

In order to gain physical insight into the loop representation, it seems
natural to apply these general ideas to simpler systems. Perhaps the
simplest toy-model is general relativity in $2+1$-dimensions. This model
has been investigated in some detail \cite{5} and has provided insights
into several ``geometric'' aspects of  the loop representation, particularly
on the role of the diffeomorphism invariance of the theory in the
quantum description.   However, in this model  there are only a finite
number of degrees of freedom, whence issues related to the presence of
the infinite dimensional reduced phase-space of $3+1$ general relativity
have remained unexplored. To gain insight into these issues, it is natural
to use the loop representation for quantization of Maxwell and Yang-Mills
theories. The goal is three-folds. First, one wants to check if one can
indeed reproduce the known, ``correct'' quantum physics via loop
quantization. A positive result would give considerable confidence in the
loop approach since nowhere in its development was it {\it required} that
it should yield, e.g., the Poincar\'e invariant Fock space of photons.
Second, one hopes that the results obtained in these model systems may
provide the much needed physical intuition for loop states, as well as
technical tools that may be useful in solving the open mathematical
problems in the gravitational case. Third, within the models themselves,
one may be able to apply the loop space methods to obtain new results.
For example, one can argue that the loop representation is especially
well-suited to the computation of higher excited states and their energy
levels in lattice QCD, to couple the theory to dynamical fermions, to
provide a firmer relation between string like structures ({\sl a la} Nambu
and Goto) and QCD, to analyze properties of coherent states in QED and
in the discussion of issues related to flux quantization in condensed
matter physics.

The purpose of this paper is to provide a loop quantization of the free
Maxwell field. In particular we will construct, entirely within the loop
picture, the analog of the Fock space of photons and represent on it
physically interesting operators, including the Hamiltonian. Furthermore,
we will provide a direct interpretation of the loop states themselves.
Finally, the mathematical techniques we introduce to obtain the
inner-product between physical states are likely to be useful in quantum
gravity.

Perhaps the most intriguing aspect of the loop representation is that the
quantum states are now functionals of loops. In field theory, one often
uses a functional representation. However, in these descriptions, the
states are functionals of the {\it dynamical variables} of the theory. For
example, in the case of the Maxwell field, one often use a basis $\mid\!
A(\vec x)\rangle$, on which the field operator $\hat{A}_a(\vec x )$ acts
as a multiplication operator, and represents states as functionals of the
vector potential $A(\vec x )$. In a loop representation, by
contrast, the states are functionals, $\Psi [\gamma ]$, of closed loops
$\gamma$ on a $3$-manifold. Now, closed loops themselves have no {\it
dynamical} significance whatsoever in the classical Maxwell theory. How
can such a description then be viable and why should it have any relation
at all to the Fock space, represented, e.g., as the space of functionals of
vector potentials?

Let us first discuss a simple example to see that there is no a priori
conflict. Consider, within the standard non-relativistic quantum mechanics,
a Hydrogen atom. Normally, its states are represented as functions,
$\Psi (\vec{x})$, of the configuration variable $\vec{x}$. However, one
can also go to the basis in which the Hamiltonian, $\hat{H}$, the total
angular momentum, $\hat{L}^2$, and the $z$-component of the angular
momentum, $\hat{L}_z$, are diagonal, and represent states as functions,
$\Psi [n,l,m]$ of three integers, $n,l,m$. The sets of these three integers
have no physical significance at all in the classical dynamics of the
hydrogen atom. And yet the representation of states as functions
$\Psi [n,l,m]$ is both viable and extremely useful. If some one hands us
just the space of (normalized) functions $\Psi [n,l,m]$, it would
be difficult to see they have anything to do what so ever with the
Coulomb problem. However, if in addition we are given the physical
interpretation of a complete set of operators $\hat{H}, \hat{L}^2,
\hat{L}_z$ which act by multiplication on $\Psi[n,l,m]$  --i.e., $\hat{H}
\cdot\Psi[n,l,m] := (-13.6 {\rm ev}/n^2)\Psi [n,l,m]$, etc-- then it
follows immediately that the functions $\Psi [n,l,m]$ of $3$ integers
represent the quantum states of the hydrogen atom. Furthermore, we can
then express any other physical observable --such as $\hat{X}$-- as an
operator on these wave functions.

The situation is similar in loop representations. In the one constructed
in this paper, it is the (positive frequency) electric fields --which
form a complete commuting set-- that act as multiplication operators
on loop states. The fact that the loops are closed ensures consistency
with the constraint that electric fields must be divergence-free. Armed
with this interpretation, not only can one relate the loop states with the
familiar Fock space but one can also express physical observables
{\it directly} in the loop representation. This detailed analysis removes
much of the unease that one experiences when one first encounters the loop
representation in canonical gravity.

Since our motivation comes from quantum gravity, the paper is
primarily addressed to gravitational theorists. Therefore, we begin
in section 2 by recalling the Bargmann representation for a quantum
Maxwell field which may be unfamiliar to our intended audience.
In section 3, we exhibit the transform from the Bargmann to the loop
representation, show that it exists rigorously and discuss its
properties. In particular, we present the loop states which correspond
to the n-photon states, normally used as the basis in the Fock description,
as well as the images in the loop representation, of the familiar operators
on the Fock space. The Bargmann states (section 2) have been used widely,
especially in quantum optics, and the general ideas behind the loop
transform (section 3) are well-known in the literature on lattice gauge
theories. The new element here is the synthesis of the two ideas. Indeed,
the loop transform in the continuum is well-defined {\it precisely} because
we depart from the practice of using wave functions $\Psi(A)$ of real
connections and use instead the Bargmann states which are holomorphic
functionals of the positive frequency connections. In section 4, we
construct the loop representation ab initio, without {\it any} reference
to the Bargmann representation or the loop transform. Strictly, from
the viewpoint of the Maxwell theory alone, this step is unnecessary.
Its motivation comes, rather, from quantum gravity
where, as noted above, the transform exists only as a formal device.
In section 5, we point out that the quantization procedure used in
section 4 is precisely the one that was first introduced in the
gravitational case and comment on the relation between our framework to
other similar treatments.

The idea of formulating quantum theory of gauge fields in  terms of
loops and holonomies has in fact a long tradition \cite{loops}. In
particular, a somewhat different treatment of the quantum Maxwell theory
in terms of loops was given by Gambini and Trias \cite{abeliangambini}.
The present paper itself belongs to a series. The second paper \cite{6} in
this series provides a general framework for lattice QCD with fermions,
the third \cite{7} uses this framework to carry out computations of
ground and excited state energies in 2+1-dimensional QCD without
fermions. Continuum QCD without fermions is discussed in \cite{qcd}. Finally.
the analogous loop representation of linearized gravity (i.e. gravitons) is
given in \cite{8}.

\section{Bargmann representation}

Our purpose here is to recall the Bargmann representation
\cite{9}  of free photons, using, however, a canonical framework. In
this representation, quantum states arise as holomorphic functionals of
complex connections.  The choice of this representation as the point of
departure is a key element of our analysis: we will see in the next section
that it is this strategy that enables us to pass to the loop picture in a
straightforward fashion, without encountering any divergences.
In addition, the general procedure used is closely related to the one
proposed \cite{2} in gravity where it is the canonical quantization
method and the holomorphic connection representation that are naturally
available.

Let us begin with a summary of the phase space description of the
Maxwell field.  Let $\Sigma$ denote a $t=$const. slice in Minkowski space.
$\Sigma$ is topologically $R^3$ and equipped with a flat, positive definite
metric $q_{ab}$.  The configuration variable for the Maxwell field is
the $U(1)-$connection 1-form $A_a\x$ on
$\Sigma$, the vector potential for the magnetic field $B^a \x$.  Its
canonically conjugate momentum is the electric field $E^a \x$. The
fundamental Poisson bracket is
\f
    \{  A_a\x,  E^b(\vec y) \}  =  \delta_a{}^b \ \delta^3(\vec x,
     \vec y).
\ff
The system has one first class constraint, $\partial_a E^a=0$. One can
therefore pass to the reduced phase space $\hat \Gamma$, by, for example,
fixing the Coulomb gauge. Let us do so. The true dynamical degrees of
freedom are then contained in the pair $ (A^T_a \x, E_T^a\x) $ of
transverse (i.e. divergence free) fields on $\Sigma$. The only non
vanishing Poisson bracket now is
\f
                \{ A_a^T \x ,  E_T^b(\vec y) \}  =
        \delta_a{}^b \ \delta^3(\vec x, \vec y)
        - \Delta^{-1} \partial_a \partial^b \, \delta^3(\vec x, \vec y),
        \label{poisson}
\ff
where $\Delta$ is the negative of the Laplacian of $q_{ab}$. (Thus $\Delta$
is a non-negative operator.)

It is easier --although, by no means essential-- to work with momentum
space variables. Let us therefore express $A_a^T\x$ and $E^a_T\x$ in
terms of their Fourier components:
\begin{eqnarray}
        A^T_a\x  &= & {1 \over  (2\pi)^{3/2}}
        \int d^3k \  e^{ i  \k \cdot \vec x}\
        [q_1(\k) m_a(\k) + q_2(\k) \bar m_a(\k)] ,
\nonumber \\
        E^a_T\x  &= & {-1\over(2\pi)^{3/2}}
        \int d^3k \  e^{ i  \k \cdot \vec x}\
        [p_1(\k) m^a(\k) + p_2(\k) \bar m^a(\k)] ,
\end{eqnarray}
where $m_a(\k)$ is a complex vector field in the momentum space which
is transverse, $m_a(\k)k^a=0$, and normalized such that
$m^a(\k)m_a(\k)= 0,\ \ \  m^a(\k)\bar m_a(\k)= 1$. (The negative sign in
front of the r.h.s of the second
equation is inserted only to simplify subsequent equations.) The
$q_j(\k)$ and $p_j(\k)$, with $j=1,2$, capture the true degrees of
freedom in the phase space of the Maxwell field;
they describe the two radiative modes of the Maxwell field
corresponding to the two helicities.
The Poisson bracket relations (\ref{poisson}) are equivalent to:
\f
        \{ q_i(-\k), p_j(\k'\,) \} =  \delta_{ij}\, \delta^3(\k,\k'\,).
\ff

In the Bargmann representation, one works with positive and negative
frequency fields. Given a pair $(A^T, E_T)$ in the reduced phase space,
one can evolve it to obtain a (real) solution to Maxwell's source-free
equations, decompose it into positive and negative frequency parts and
consider the data $(\ppmm A , \ppmm E)$ induced on the initial slice by
the two parts. It is easy to verify that the {\it positive frequency\ }
connection $\p A$ is given by
\begin{eqnarray}
        \p A_a\x & = &  {1\over\sqrt{2}}
        \left(A^T_a\x +  i \,   \Delta^{- 1/2}   (E_{T})_a\x\right)
\nonumber \\
        & \equiv &  {1\over (2\pi)^{3/2}}    \int {d^3\k\over|\k|}\
        e^{ i  \k \cdot \vec x}\
        \left[\zeta_1(\k) m_a(\k) + \zeta_2(\k) \bar m_a(\k) \right] ,
        \label{fouriercomp}
\end{eqnarray}
with
\f
         \zeta_j(\k) =
          {1\over\sqrt{2}}\left( |\k| q_j(\k) -  i \, p_j(\k) \right).
\ff
It is conjugate to the {\it negative frequency}  electric field,
$ \m E$  given by
\begin{eqnarray}
                \m E^a\x & = &{1\over\sqrt{2}}
                \left(E_T^a\x +  i  \Delta^{1/2} (A^{T})^a\x\right)
\nonumber \\
                & \equiv &{-i\over (2\pi)^{3/2}}
        \int {d^3\k}\      e^{ i  \k \cdot \vec x}\
        \left[
        \bar \zeta_1(-\k) m_a(\k) + \bar \zeta_2(-\k) \bar m_a(\k)
        \right] ,
        \label{fouriercomp2}
\end{eqnarray}

The true degrees of freedom are now coded  in the two {\it complex}
fields $ \zeta_j(\k) $ in momentum space: the Fourier coefficients of
the positive frequency, transverse connection $ \p A_a\x $. These
$ \zeta_j(\k) $ provide us with a complex  chart on the reduced phase
space of the real Maxwell field.   The only non-vanishing Poisson brackets
among $ \zeta_j(\k) $ and $ \bar \zeta_j(\k)$ are:
\f
            \{  \zeta_i(\k), \  \bar\zeta_j(\k') \}  =
            i |\k|\  \delta_{ij} \, \delta^3(\k,\k').
\ff
The $\zeta_j(\k) $ play the same role as the complex coordinates, $ z =
{1\over\sqrt{2}} (\omega q -  i  p) $ on the phase spare of a
harmonic oscillator.  Thus, the Maxwell field is now represented as an
assembly of harmonic oscillators, two for each 3-momentum $ \k $.

To  quantize this system, let us begin by letting the $ \zeta_j(\k) $ and
$\bar \zeta_j(\k)$ be the ``elementary classical variables'' which are to
have unambiguous quantum analogs, $ \hat \zeta_j(\k) $ and
$ \hat{\bar\zeta}_j(\k) $. To begin with, we consider them as abstract
operators. Let $\cal A $ denote the associative algebra
they generate, subject to the canonical commutation relations (CCRs):
\f
            [  \hat \zeta_i(\k), \hat{\bar\zeta}_j(\k') ]  =
           - \hbar |\k|\  \delta_{ij}\, \delta^3(\k,\k').
            \label{CCRs}
\ff
On this algebra, we define an involution operation, $*$, via
\f
          \left(  \hat \zeta_i(\k)  \right)^* =  \hat{\bar\zeta}_j(\k).
           \label{star}
\ff
Denote the resulting $*$-algebra by ${\cal A}^{(*)}$.  In the quantum
theory, the abstract operators that form ${\cal A}^{(*)}$ are realized as
linear operators on a Hilbert space of states. Thus, we now construct the
quantum theory as a linear representation of the $*$-algebra ${\cal
A}^{(*)}$ on a Hilbert space.

In the Bargmann method, this representation is constructed by using the
``anti-holomorphic polarization'' on the reduced phase space. Thus, one
begins with the vector space $V$ of functionals $\Psi(\zeta_j)$ which
are polynomials in $\zeta_j(\k) $.  (Since they are independent of $ \bar
\zeta_j(\k)$, they are entire, holomorphic.) \  On this $ V$, the operators
$\hat \zeta_j(\k) $ are introduced in the obvious fashion
\f
        \hat\zeta_j(\k)  \circ \Psi(\zeta_i) =  \zeta_j(\k) \Psi(\zeta_i) .
        \label{zeta}
\ff
To ensure that the CCRs (\ref{CCRs}) are satisfied, it is then natural to
represent $ \hat{\bar\zeta}_j(\k)  $ via

\f
       \hat{\bar\zeta}_j(\k)  \circ \Psi(\zeta_i)  =
        \hbar |\k| \  {\delta \Psi(\zeta_i)\over\delta\zeta_j(\k)}.
        \label{zetabar}
\ff
The final task is to equip $ V $ with an inner-product. The idea \cite{11}
is to use the $*$-relations (\ref{star}) to constraint the inner product.
Thus, we seek an inner product $ <\ ,\ >$ with respect to which the
operator appearing on the right side of (\ref{zetabar}) is the
Hermitian adjoint of that appearing on the right side of  (\ref{zeta}),
i.e. such that
\f
      < \Psi,\  \zeta_j(\k)\Phi > \ \  = \ \
      < \hbar |\k|  {\delta\Psi\over\delta\zeta_j(\k)},\  \Phi >,
      \label{quantrealcond}
\ff
for all $ \Psi $ and $\Phi$ in $ V $. Since the $*$-relations (\ref{star})
capture, in the quantum theory, the fact that $(A^T, E_T ) $ are real
classical fields, (\ref{quantrealcond}) is equivalent to demanding that
the quantum field operators  $(\hat A^T, \hat E_T )$ be self- adjoint.
Hence the requirement (\ref{quantrealcond}) on the permissible inner
products will be called the {\it quantum reality condition}.

To find an inner product that satisfies this condition, we utilize
a formal procedure; the resulting inner product, however, will
be well defined. We begin with the ansatz
\f
         <\Psi, \Phi>  =  \int_{\hat\Gamma} \prod_j
        {\rm d\!I}\zeta_j \wedge {\rm d\!I} \bar\zeta_j
         \mu(\zeta_j ,\bar\zeta_j)  \overline{\Psi(\zeta_j)}
         \Phi(\zeta_j),
\ff
where $ \prod_j {\rm d\!I}\zeta_j \wedge {\rm d\!I}\bar\zeta_j $
is the `` translation invariant volume element''
on the reduced phase space $\hat\Gamma$  and regard
(\ref{quantrealcond}) as a condition on the ``measure'', namely on
$\mu(\zeta_j, \bar\zeta_j)$. Using the definitions (\ref{zeta},
\ref{zetabar}) of  $\hat\zeta_j(\k)$ and $\hat{\bar\zeta}_j(\k)$, it
is straightforward to verify \footnote {In this paper we ignore the
issues of domains of various (densely defined) operators. In particular,
we do not distinguish between self-adjoint and symmetric operators.} that
(\ref{quantrealcond}) determines $\mu(\zeta_j ,\bar\zeta_j)$ uniquely
(except for a trivial multiplicative constant) as
\f
          \mu(\zeta_j ,\bar\zeta_j) = \exp\ \  - \int {d^3\k\over{\hbar|\k|}}\
        (|\zeta_1(\k)|^2  + |\zeta_2(\k)|^2).
        \label{measure}
\ff
Thus, $\prod_j  {\rm d\!I}\zeta_j \wedge {\rm d\!I}\bar\zeta_j\
\mu(\zeta_j,  \bar\zeta_j)$  turns out to be a (well-defined) Gaussian
measure.
The Hilbert space $ \cal H $ of all quantum states is obtained by the
Cauchy completion of $ V $ with respect to this inner product.  Note that
the quantum reality conditions have led us directly to the standard
Gaussian measure; we did not have to invoke Poincar\'e invariance of the
inner product or of the vacuum. One may therefore expect this procedure
to select the inner product to succeed also beyond Minkowskian field
theories. This expectation is borne out in, e.g., 2+1-dimensional gravity.
\cite{5}

We conclude this section by exhibiting a few states and operators in the
Bargmann representation. The Fock  vacuum  $ |0>$ is represented by the
functional $ \Psi(\zeta_j) = 1 $. The 1-photon state, $| \k_0, \epsilon=1>$,
with momentum $\k_0$ and helicity $ +1$, is represented by the
linear functional $ \Psi(\zeta_j) = \zeta_1(\k_0) $ which assigns to each
$ \zeta_j(\k) $ the value of $\zeta_1 $ at $ \k_0$.  A general 1-photon
state, say with helicity $ +1 $, is a superposition,
$\int {d^3\k\over{\hbar|\k|}} \bar f(\k) | \k, \epsilon=1 >$, of the pure
momentum  states where $\int {d^3\k\over{\hbar|\k|}} |f(\k)|^2<\infty $.
In the Bargmann picture it is represented by the functional $ \Psi(\zeta_j)
= \int {d^3\k\over {\hbar|\k|}} \bar{f}(\k)  \zeta_1(\k) $. A n-photon state
is represented by a sum of
products of 1-photon wave functionals:
\f
        \Psi_n(\zeta_j) =
        \int {d^3\k_1\over{\hbar|k_1|}}...\int {d^3\k_n\over{\hbar|k_n|}}
       \bar{f}_{j_1...j_n}(\k_1,...,\k_n)\
        \zeta_{j_1}(\k_1) ... \zeta_{j_n}(\k_n),
        \label{nphotons}
\ff
where the repeated indices $j_i$ are summed over $j_i= 1,2$. The operators
$\hat\zeta_j(\k)$ are the creators; $ \hat{\bar\zeta}_j(\k)$ are the
annihilators.   Finally, the normal ordered  Hamiltonian is given by
\f
        : \hat H : \ \ = \int d^3\k\ \hbar |\k|\
        \sum_{j=1}^2 \ \zeta_j(\k) \ {\delta\over\delta\zeta_j(\k)}.
\ff

This completes the summary of the Bargmann representation of the free
Maxwell field.

\section{A loop transform}

The idea now is to pass to a loop representation by performing a
transform from functionals $\Psi({}^+\!A) \sim \Psi(\zeta_j)$ of
transverse, positive frequency connections $ \p A_a\x $ on $\Sigma$
via equation (\ref{fouriercomp})\ ) to functionals
$ \psi(\gamma) $ of closed loops $\gamma $ on $\Sigma $.
(Recall that $ \p A_a\x \equiv (\zeta_1(\k), \zeta_2(\k).)$ The
transform is analogous to the Fourier transform
\f
      \psi(\vec p) = {1\over (2\pi)^{3/2}}
      \int d^3\vec x \  e^{ i \vec p \cdot \vec x}\ \Psi\x,
\ff
which enables one to pass from the position to the momentum
representation in non-relativistic quantum mechanics. The idea
underlying this loop transform was first introduced by Rovelli and Smolin
\cite{2} in the context of canonical gravity, where they used the
expression
\f
       \psi(\gamma) = \int d\mu(A)\ \overline{T[\gamma,A]} \ \Psi(A),
        \label{transform}
\ff
as a formal tool to pass from the connection representation, in which
states $\Psi(A)$ arise as holomorphic functionals of (self dual)
connections, to the loop representation, in which states are certain
functionals on the loop space.   Here, the trace of the holonomy
$T[\gamma,A]$ of the gravitational connection $A_a^i\x$ around the
closed loop  $\gamma$ is analogous to the Kernel $ e^{ i  \k \cdot \vec x}
$ of the  Fourier transform and $d\mu(A)$ is to be a suitable measure on
the  space of connections.  However, as pointed out already in \cite{2},
equation  (\ref{transform}) is only a formal expression because we do not
yet know  an appropriate measure on the space of gravitational
connections  $ A_a^i\x $. \cite{4}\ \   We will see in this section that the
transform {\it  can\,}, however, be made rigorous in the case of the
Maxwell field:  not only does $\psi(\gamma) $ exist for all Bargmann
states $\Psi(\zeta_j)$,  but it can also be evaluated explicitly !

This section is divided into three parts. In the first, we introduce some
notions associated with closed loops on $\Sigma$, in the second we show
that the transform exists and in the third we show that it is faithful.

\subsection{Loops}

Let us begin with some definitions. By a {\it loop\,} we shall mean a
continuous and
piecewise smooth mapping $\gamma(s)$ from $S^1 $ to $ \Sigma$, where
$ s\in[0,2\pi[$ (the end points $ 0$ and $2\pi $ being identified). Two
loops $\gamma$ and $\beta$ will be said to be {\it holonomically
equivalent\ } if, for every smooth connection $A_a$, we have $
\oint_\gamma A_a ds^a =  \oint_\beta A_a ds^a$.  Thus, if holonomically
equivalent, $\gamma$ and $\beta$ can differ from each other only
through

\noindent i) reparametrization, $\gamma(s) = \beta(s')$ for some
reparametrization $ s \rightarrow s' $ of the curve $ \beta(s)$.
(Note that the reparametrization does not have to be continuous at the
points where $\beta$ intersects itself);

\noindent ii) retracing identity, $ \gamma = l \circ \beta \circ l^{-1}$,
where $l$ is a line segment and $\circ$  indicates the obvious
composition of segments;

\noindent or, any combination of retracings and reparametrizations.

\noindent Each equivalence class will be referred to as a
{\it holonomic loop\ }  and the set of all these equivalence classes
will be called the {\it holonomic loop space\ } and denoted by $\cal HL$.

Given a loop $\gamma$, we define its {\it form factor},
$F^a(\gamma, \vec x)$, to be a distributional vector density of weight
one via:
\f
           \int F^a(\gamma, \vec x) \ \omega_a \x\  d^3{\vec x} =
           \oint_\gamma  \omega_a \  ds^a .
           \label{formfactor}
\ff
Thus, $F^a(\gamma, \vec x)$ may be more directly expressed as
\f
             F^a(\gamma, \vec x) = \oint_\gamma ds \
             \dot\gamma^a(s)\ \delta^3(\vec x, \vec \gamma(s)),
\ff
where $\vec\gamma(s)$ is the point on the loop $\gamma$ at parameter
value $s$ and $\dot\gamma^{a}(s)$ the tangent vector to $\gamma$ at
$\vec\gamma(s)$.  It follows immediately from the definition
(\ref{formfactor}) that
\f
            \int F^a(\gamma, \vec x) \ \partial_a\omega\x \ d^3x = 0
\ff
for all $\omega\x$, whence $ F^a(\gamma, \vec x) $ is divergence-free:
\f
             \partial_a F^a(\gamma, \vec x) = 0.
\ff
Note, incidentally, that neither the notion of the form factor nor its
properties refer to the metric or even the topology of $ \Sigma $.
Therefore the notion is useful well beyond Maxwell theory. \cite{13}\ \
For the purpose of Maxwell theory, however, it is convenient to use the
flat metric $q_{ab}$ and perform a Fourier transform to obtain the
momentum space representation of $ F^a(\gamma, \vec x) $. We have:
\begin{eqnarray}
          F^a(\gamma, \k) & \equiv & {1\over(2\pi)^{3/2}}
          \int d^3x \ e^{-i\k \cdot \vec x} \ F^a(\gamma,\vec x)
   \nonumber \\
        & = & {1\over(2\pi)^{3/2}} \oint_\gamma ds \ \dot\gamma^a(s)\
        e^{-i \k \cdot \vec \gamma(s)} .
\end{eqnarray}

Let us note a few properties of these form factors.  First, it follows
from the very definition (\ref{formfactor}) that two loops $\gamma$ and
$\delta$ have the same form factors if and only if they are holonomic.
Thus, $F^a(\gamma,\vec x)$ --or $ F^a(\gamma,\k)$-- can be used to
characterize holonomic loops, whence the name ``form factors''. (Note,
however, that they do {\it not\ } coordinatize $\cal HL$ in the sense
normally used in differential geometry because, given a loop $\gamma$
with the form factor $F^a(\gamma,\k)$, there is in general no loop
$\delta$ such that its form factor satisfies $F^a(\delta,\k) = \lambda
F^a(\gamma,\k)$ for a general real number $\lambda$.  Hence,
attractive as it may first seem, the strategy of using form factors to
induce a manifold structure on $\cal HL$ fails to work.)
Next, since $F^a(\gamma,\vec x)$ is divergence-free,  $F^a(\gamma,\k)$ is
transverse:  $k_a F^a(\gamma,\k)= 0$. Hence it has only two independent
components.  Let us label them by  $F_j(\gamma,\k)$:
\begin{eqnarray}
          F_1(\gamma, \k) & = & {\hbar\over(2\pi)^{3/2}}
        \oint_\gamma ds \ \dot\gamma^a(s)\   \bar m_a(\k)\
        e^{-i  \k \cdot \vec \gamma(s)} ,
  \\
        F_2(\gamma, \k) & = & {\hbar\over(2\pi)^{3/2}}
        \oint_\gamma ds \ \dot\gamma^a(s)\   m_a(\k)\
        e^{-i  \k \cdot \vec \gamma(s)} ,
\end{eqnarray}
so that
\f
        F^a(\gamma, \k) =  {1\over \hbar} \
        (F_1(\gamma, \k) m^a(\k) + F_2(\gamma, \k) \bar m^a(\k)).
\ff
Here, we have chosen normalization and other conventions that will simplify
later calculations. This transversality of form factors will play an important
role in the loop quantization because it captures in a natural way the gauge
invariance of the theory, i.e., the transversality of the photon. The next
property follows from the fact that $ F^a(\gamma, \vec x) $ is real; its
Fourier transform $ F^a(\gamma, \k) $ sa\-tisfies the ``reality condition''
$\bar F_j(\gamma, \k) = -  F_j(\gamma, -\k) $. (The fields $q_j(\k)$ and
$p_j(\k)$ introduced in section 2 satisfy the same ``reality conditions'':
$\bar q_j(\k) = - q_j(-\k)$ and $\bar p_j(\k) = - p_j(-\k)$. The fields
$\zeta_j(\k)$, on the other hand, are Fourier coefficients of a complex field,
$\p A_a^T\x$, and are therefore not subject to these restrictions).

Finally, given two holonomic loops $\gamma$ and $\delta$, we define a new
holonomic loop, $\gamma \# \delta$, called the {\it eye-glass loop\ } as
follows: $\gamma \# \delta \equiv l \circ \gamma \circ l^{-1} \circ \delta$
where $l$ is any line segment  joining a point on $\gamma$ to a point on
$\delta$. \footnote{Note that the operation on the loop space defined here
by $\#$ is different from the one defined in the non-Abelian case.
However, the difference arises only because of the trace identities for
$U(1)$ and $SL(2, C)$ are different. There is a well-defined sense
in which the operation defined here is the Abelian analog of that defined
in \cite{2}.}

Thus, if $p$ and $q$ are points on loops $\gamma$ and $\delta$
respectively, and $l$ joins $p$ to $q$, then $\gamma \# \delta$ is
the holonomic loop obtained by first going around $\gamma$ starting and
ending at $p$, then going along $l$ from $p$ to $q$, then around $\delta$,
and then back along $l$ to the point $p$ on $\gamma$. Although the
specific loop so obtained depends on the choices of $p,q$ and $l$, the
resulting holonomic loop is the same. Using the definition of form factors,
we now have:
\f
           F_j(\gamma\#\delta, \k) = F_j(\gamma,\k) + F_j(\delta,\k).
\ff
In particular, if $\gamma = \delta$, we have $ F_j(\gamma\#\gamma,\k)
= 2 F_j(\gamma,\k) $. Thus, the space of form factors -- i.e., the space
of the fields $F^a(\k)$ such that there is a loop $\gamma$ for which
$F^a(\k)=F^a(\gamma, \k)$ \ )-- does have the structure of a vector
space over, however, the ring of integers rather than the ring of reals or
complexes.

\subsection{A transform}

We are now ready to define a loop transform. As remarked at the
beginning of this section, to define the analog of (\ref{transform}),
we need, for Maxwell fields, the analog of the trace of the holonomy
and a measure on the space of positive frequency connections $\p A_a\x$.
The analog of $T[\gamma, A]$ is simply $T[\gamma, \p A] = \exp
\oint_\gamma \p A_a ds^a$. (We have set the electric chanrge equal to 1
and omitted the conventional factor of $i$ in front of the integrand
because now the connection $\p A_a$ is itself complex). For the measure,
we do have a satisfactory candidate: the Poincar\'e invariant measure
(\ref{measure}) that defines the inner product in the Bargmann
representation. Let us therefore set
\f
        \psi(\gamma) =
        \int\prod_j {\rm d\!I}\zeta_j\wedge {\rm d\!I}\bar\zeta_j\
        e^{-\int {d^3\k\over{\hbar|\k|}}|\zeta_j(\k)|^2}\ \
        e^{\overline{\oint \p A_a ds^a}} \ \ \Psi(\zeta_j).
        \label{abtransf}
\ff
The question is if the integral exists for all Bargmann states
$\Psi(\zeta_j)$ and, if so, whether we can evaluate it explicitly. We
shall show that the answer to both questions is in the affirmative.

Let us begin by re-expressing the holonomy in terms of $\zeta_j$. Using
the expression (\ref{fouriercomp}) of $\p A\x$ in terms of its Fourier
components, we have:
\begin{eqnarray}
        \oint \p A_a \ ds^a & = &
        {1\over(2\pi)^{3/2}}  \int {d^3\k\over|\k|}
        \left[
        \zeta_1(\k)      \oint_\gamma
        e^{ i  \k \cdot \vec \gamma(s)}
        \dot\gamma^a(s)    m_a(\k) ds
        \right.
\nonumber  \\   &  &
        \left.
         +\ \zeta_2(\k) \oint_\gamma
        e^{ i \k\cdot\vec\gamma(s)}
        \dot\gamma^a(s) \bar m_a(\k) ds
        \right]
\nonumber \\
        & = & \int {d^3\k\over {\hbar|\k|}} \
                \sum_j \ \zeta_j(\k)\ \bar F_j(\gamma,\k).
        \label{holonomy}
\end{eqnarray}

Thus, using the form factors, we have re-expressed the holonomy as an
integral in momentum space, where the integrand splits up into two
factors; the first, $\zeta_j(\k)$, depending only on the connection
$\p A\x$ and the second, $F_j(\gamma,\k)$, depending only on
the the holonomic loop $\gamma$.  Let us substitute (\ref{holonomy})
into (\ref{abtransf}). The question then reduces to that of the existence
(and of the value) of the Gaussian integral
\f
        \psi(\gamma) =
        \int\prod_j {\rm d\!I}\zeta_j\wedge {\rm d\!I}\bar\zeta_j\ \
        e^{-\int {d^3\k\over {\hbar|\k|}}|\zeta_j(\k)|^2}\ \
        e^{ \int {d^3\k\over {\hbar|\k|}} \bar\zeta_j(\k)F_j(\gamma,\k) }
        \ \ \Psi(\zeta_j).
        \label{integral}
\ff

Let us begin by focussing on just one of the doubly infinite modes
contained in $\zeta_j(\k)$.  If we denote this mode simply by $\zeta$,
the analog of (\ref{integral}) is
\f
        \psi(F) = \int {d\zeta\wedge d\bar\zeta\over 4\pi i \hbar}
        \ e^{-\zeta \bar\zeta} \ e^{-F \bar\zeta} \ \Psi(\zeta).
        \label{analog}
\ff
where $\Psi(\zeta)$ is a Bargmann state for a single harmonic oscillator
representing the mode $\zeta$. Let us therefore ask if the integral on the
right side of (\ref{analog}) exists for all elements $\Psi(\zeta)$ of the
Bargmann Hilbert space. Set
\f
        C_F(\zeta) = \exp \bar F\zeta .
\ff
Then the integral in question is precisely the Bargmann inner product
between the states $C_F(\zeta)$ and $\Psi(\zeta)$ :
\f
        \psi(F) = <C_F,\ \Psi>.
        \label{innerprod}
\ff
Note that $C_F(\zeta)$ is precisely a coherent state in the Bargmann
Hilbert space. (It is normalizable, with norm $<C_F,\ C_F> = \exp F
\bar F$.)
Thus, not only does the inner product  (\ref{innerprod}) exist for all
Bargmann states $\Psi(\zeta)$, but the result, $\psi(F)$, can be
interpreted as giving the components of $\Psi$ along the (overcomplete)
{\it coherent state basis}.  Furthermore, the explicit expression of
$\psi(\zeta)$ is easy to evaluate. Let $ \Psi(\zeta) = \sum a_n\zeta^n $.
Expanding the coherent state  $C_F(\zeta)$ in a power series and using
the fact that, being eigenstates of the number operator, the functions
${1\over\sqrt{n!}}\zeta^n$  are orthonormal, we find  $ \psi(F) = \sum
a_n F^n$ . Since the polynomials are dense in the Bargmann Hilbert space,
it follows that the image  $\psi(F)$  of the Bargmann state  $\Psi(\zeta)$
under the transform  (\ref{analog})  is simply
\f
        \psi(F) = \Psi(F) .
\ff
The transform is astonishingly simple precisely because of the use of
the coherent state basis, i.e. of the Bargmann representation.

Let us now return to the doubly infinite modes,  $\zeta_j(\k)$ of the
Maxwell field. The analogs of the polynomials  $\sum a_n \zeta^n$  are
the cylindrical functionals $\Psi(\zeta_j) = \sum_n\int {d^3\k_1\over
{\hbar|\k_1|}}... {d^3\k_n\over{\hbar|\k_n|}}\  \bar
a^{(n)}_{j_1...j_n}(\k_1,...,\k_n)\ \zeta_{j_1}(\k_1) \zeta_{j_n}(\k_n)\ $,
where the coefficients  $a^{(n)}_{j_1...j_n}(\k_1,...,\k_n)$  are such
that the integral
$ \int {d^3\k_1\over{\hbar|\k_1|}}  ... \int {d^3\k_n\over{\hbar|\k_n|}}|
a^{(n)}_{j_1...j_n}(\k_1,...,\k_n)|^2$
converges. These cylindrical functionals are well-defined on the 1-photon
Hilbert space  $H_1$  spanned by the  $\zeta_j(\k)$  for which  $ \int
{d^3\k\over{\hbar|\k|}} (|\zeta_1(\k)|^2+|\zeta_2(\k)|^2) $  converges.
This is precisely the Hilbert space on which the integral in the loop
transform (\ref{abtransf}) is being defined. From the definition of
Gaussian integrals on Hilbert spaces and the result discussed above for
a harmonic oscillator, it now follows that the loop transform exists for
all Bargmann photon states  $\Psi(\zeta_j)$. Furthermore if we denote the
transform by  $\cal T$ ,
\f
        {\cal T} \circ \Psi(\zeta_j) = \psi(\gamma),
\ff
where  $ \psi(\gamma) $  is given by  (\ref{abtransf}), we have the
simple result:
\f
        {\cal T} \circ \Psi(\zeta_j) = \Psi(F_j(\gamma)),
\ff

Thus, the transform does map Bargmann states to well defined
functionals on the holonomic loop space  $\cal HL$  which depend on
holonomic loop $\gamma$ only through their form factors
$F_j(\gamma,\k)$.

Let us consider a few examples. Since the Fock vacuum $|0>$ is
represented by the functional $\Psi_0(\zeta_j) =1 $ in the Bargmann
representation, it is represented by the unit functional,
$ \psi_0(\gamma) =1 $ also in the loop representation.
A one photon state,  $ |\k_0,\epsilon=1>$, with
momentum  $\k_0$  and helicity $+1$, is represented by
$\Psi_{k_0,+1}(\zeta_j) = \zeta_1(\k_0)$ in the Bargmann
representation and hence by the functional
\begin{eqnarray}
        \psi_{k_0,+1}(\gamma) & = & F_1(\gamma,\k_0)
\nonumber \\
        & = & {\hbar\over(2\pi)^{3/2}}\oint_\gamma ds\ \
        \dot\gamma^a(s) e^{-i\k_0\cdot\vec\gamma(s)} \bar m_a(\k_0)
\nonumber \\
        & = & \hbar \oint_\gamma ds^a \bar{A}_a^{(\k_0,+1)}
\end{eqnarray}
in the loop representation, where $A_a^{(\k_0,+1)}= {1\over(2\pi)^{3/2}}
e^{i \k_0\cdot\vec\gamma(s)} \  m_a(\k_0)$ is the plane wave
connection with momentum $\hbar \k_0$  and helicity $+1$. A general
1-photon state is given by the superposition  $ \int
{d^3\k\over{\hbar|\k|}} f_j(\k) \ |\k,j>$. In the loop representation it
is simply
\begin{eqnarray}
        \psi_{f}(\gamma) & = & \int {d^3\k\over{\hbar|\k|}}\
        \overline{f_j(\k)}  F_j(\gamma,\k)
\nonumber \\
        & = & \oint_\gamma ds^a \bar A_a^{(f)},
        \label{Aaf}
\end{eqnarray}
where
\f
        A_a^{(f)} = {1\over(2\pi)^{3/2}} \int {d^3\k\over{\hbar|\k|}}
        e^{i \k\cdot\vec\gamma(s)}
        (f_1(\k) m_a(\k)+f_2(\k) \bar m_a(\k))
        \label{Aaff}
\ff
is the connection on $\Sigma$ defined by the given 1-photon state.
Thus, in the loop representation, 1-photon states arise simply as the
line integral of the (negative frequency) connection along the loop!
More precisely, we have the following: a normalizable,
transverse, positive frequency connection  $\p A_a\x$  gives rise
to a one photon state in the Fock space which, in the loop picture, is
represented by the line-integral of the complex conjugate of the given
connection.

The description of a n-photon state is straightforward. Recall that
in the Bargmann representation such a state is given by an n-th order
functional, $\Psi_n(\zeta_j)$, on the 1-photon Hilbert space $H_1$ (see
equation (\ref{nphotons})\ ).  Hence, its image under the transform $\cal
T$ is simply
\f
        \psi_n(\gamma) =
        \oint_\gamma ds^{a_1}...\oint_\gamma ds^{a_n}\ \
        \bar A^{(f)}_{a_1,...a_n},
        \label{nstate}
\ff
where we have used the standard (differential geometry) notation
\[
        \oint_{\gamma_1} ds^{a_1}...\oint_{\gamma_n} ds^{a_n}\
        \omega_{a_1...a_n} =
\] \f
        \oint_{\gamma_1} ds_1 \ \dot\gamma_1^{a_1}(s_1) \ ...
        \oint_{\gamma_n} ds_n \ \dot\gamma_n^{a_n}(s_n)\ \
        \omega_{a_1...a_n}(\vec\gamma_1(s_1),...,\vec \gamma_n(s_n)).
        \label{notation}
\ff
In equation (\ref{nstate}), $A^{(f)}_{a_1,...a_n}$ is the totally
symmetric field (with index $a_j$ in the tangent space of the point
$\vec x_j$ ), divergence-free in each index, given by:
\[
        A^{(f)}_{a_1,...a_n}(\vec x_1,...,\vec x_n) =
\] \[
        {1\over(2\pi)^{3n/2}}
        \int {d^3\k_1\over{\hbar|k_1|}}...\int {d^3\k_n\over{\hbar|k_n|}}\
        e^{i\k_1\cdot\vec x_1} p_{a_1}^{(j_1)}(\k_1) \ ...\
        e^{i\k_n\cdot\vec x_n} p_{a_n}^{(j_n)}(\k_n)
\] \f
        f_{j_1...j_n}(\k_1,...\k_n),
        \label{Af}
\ff
where the polarization vectors $p_{a}^{(j)}(\k) $ are given by $
p_{a}^{(1)}(\k) = m_{a}(\k)$ and $p_{a}^{(2)}(\k) = \bar
m_{a}(\k)$. Thus, an n-photon state $\psi_n(\gamma)$ is given
simply by a sum of products (\ref{nstate}) of holonomies of (negative
frequency) connections around the loop $\gamma$.

Alternatively, since $\psi_n(\gamma)= {\cal T}\circ\Psi_n(\zeta_j)=
\Psi_n(F_j(\gamma))$, equation ({\ref{Af})  can be regarded  as an
n-nomial in the form factors $F_j(\gamma,\k)$ of the loop  $\gamma$.
Although these two  descriptions of loop states are clearly equivalent,
their emphasis is  different. The first description will
be at forefront when we construct, in section 4, the loop representation
directly, without any reference to the  transform. The second, on the
other hand, is useful in translating notions and formulae between the
loop and the Bargmann pictures.

We conclude this  sub-section by pointing out a subtle difference
between the transform (\ref{analog}) for the harmonic oscillator and
the transform (\ref{abtransf}) for the Maxwell field. As pointed out
above, the integral on the right side of (\ref{analog}) can
be regarded simply as the scalar product  $ <C_F,\ \Psi>$
between the coherent state  $ C_F(\zeta) := \exp \bar F \zeta$
and the given state  $ \Psi(\zeta_j)$. Similarly, in the case of the
Maxwell field, the integral in  (\ref{abtransf})   can be
considered as the scalar product between the ``coherent state''
\f
        C_{F(\gamma)}(\zeta_j)
        := \exp\int {d^3\k\over{\hbar|\k|}}
        \bar F_j(\gamma,\k) \zeta_j(\k)
        \label{loopstate}
\ff
and the Bargmann state $\Psi(\zeta_j) $. However, now, this
interpretation is somewhat formal because the ``coherent state''
$ C_{F(\gamma)}(\zeta_j) $ is not normalizable, i.e., does not belong
to the Bargmann Hilbert space. The failure, however, is mild. To see
this, recall first that there is a  1-1 correspondence between
(normalizable) 1-photon states and coherent states, the latter being
simply the exponentials of the former. The states $C_{F_j}(\zeta_j)$
can thus be regarded as the coherent states associated with the
Bargmann 1-photon states  $ \Psi_\gamma(\zeta_j) = \int {d^3\k\over
{\hbar|\k|}}\ \ \bar{F}_j(\gamma,\k)\zeta_j(\k) $. In the physical
space picture, this is the state defined by the positive frequency
connection whose restriction to  $\Sigma$  is given by $ \p A_a\x =
F_a(\gamma,\vec x) $. Since $F_a(\gamma,\vec x) $ is distributional
(see section 3.1), the 1-photon state $\Psi_\gamma $ fails to be
normalizable.  However, its ``inner product'' with every normalizable,
1-photon state  $ \tilde\Psi(\zeta_j)$ {\it is\ } finite, being the
holonomy of the connection defined by  $ \tilde\Psi(\zeta_j)$ around
the loop $\gamma$. Furthermore, $ <\Psi_\gamma,\ \tilde\Psi> = 0 $
for all $\gamma$ if and only if the normalizable 1-photon state
$\tilde\Psi$  itself vanishes. In this sense $ \Psi_\gamma $ are
(over)complete in the 1-photon Hilbert space. Thus, the states
$\Psi_\gamma(\zeta_j) $  are analogous to the plane-wave states
$\Psi(\vec x) = \exp i\k\cdot\vec x $  which fail to be
normalizable but have finite ``inner product'' with every square
integrable function and form a complete basis. In this sense the
non-normalizability of the states  $ \Psi_\gamma(\zeta_j) $  is ``mild''.
Therefore, we can still regard  $ C_{F(\gamma)} $ defined in
(\ref{loopstate}) as a coherent state which,
however, is the exponential of the generalized 1-particle state
$\Psi_\gamma(\zeta_j)$. (The states $C_{F(\gamma )}$ are
over-complete in the {\it Fock space}.) In this sense, we can regard the
the integral in the Maxwell loop transform (\ref{abtransf}) as a scalar
product $\langle C_{F(\gamma)}, \Psi\rangle$ between generalized coherent
states $C_{F(\gamma )}$ and the given quantum state $\Psi$. In the Dirac
bra-ket notation, then, we can denote the coherent state $C_{F(\gamma )}$
simply as $|\gamma\rangle$ and the loop transform
(\ref{abtransf}) simply by
\f
        \psi(\gamma)\ \ =\ \ \langle\gamma|\Psi\rangle.
\ff

Finally, note that, both for the harmonic oscillator and the
Maxwell field, there are technical differences between the use
of an overcomplete coherent state basis and of an orthonormal basis.
For the harmonic oscillator, for example, the basis $|x\rangle$ is
orthonormal. It is an eigenbasis of the (Hermitian) position operator
$\hat{X}$ which acts by multiplication on wave functions $\Psi(x)
:= \langle x |\Psi\rangle$. The basis $|z\rangle$ (or, $|F\rangle$), on
the other hand is overcomplete and therefore {\it fails to be orthonormal}.
It is an eigenbasis of the (non-Hermitian) annihilation operator
$\hat{\bar{z}}\equiv d/dz$, but it is the operator $\hat{z}$ that acts
by multiplication on wave functions $\Psi(z) := \langle z|\Psi\rangle$.

\subsection{Properties}

In this subsection we first show that the loop transform $\cal T $
is faithful and then use this property to pull-back physically
interesting operators from the Bargmann to the loop representation.

We saw in the last subsection that the action of  $ \cal T$ is rather
simple: $\psi(\gamma) \equiv {\cal T} \circ \Psi(\zeta_j)$ is given by
$\psi(\gamma) = \Psi(F_j(\gamma)) $. Since each holonomic loop is
completely characterized by its form factor, the map $\cal T$ from the
Bargmann states to functionals on the loop space is clearly well-defined
and linear. However, it is not obvious that it is  1-1 . In the case of the
harmonic oscillator, the transform is clearly  1-1  since it maps the
holomorphic function  $ \Psi(\zeta) $  of a complex variable  $ z $  to the
holomorphic function  $ \Psi(F)  $  of a complex variable  $ F $ . In the
case of the Maxwell field, on the other hand, the argument of the
Bargmann states is two complex functions $ \zeta_j(\k) $ of three real
variables while the argument of the loop states is  (equivalence classes
at) only three real functions  $ \gamma^a(s) $  of a single real variable.
Put differently, while  $ \zeta_j(\k) $  form a complex  vector space, the
form factors  $ F_j(\gamma,\k) $  which characterize holonomic loops
constitute only a ``sparse subset'' of this vector space.\footnote{More
precisely, the situation is the following. The form factors
$F_j(\gamma,\k)$ belong to the distributional dual of the smooth
$\zeta_j(\k)$ of compact support. The Bargmann Hilbert space is, strictly
speaking, the space of weave functionals on this dual because the
Gaussian measure is concentrated on distributions. The set of form factors
is sparse in this dual.} Therefore, there does exist a large class of
functionals of  $ \zeta_j(\k) $
whose restriction to the subset of form factors vanishes identically. We
must show that none of the Bargmann states belong to this class. Roughly
speaking, the image  $ \psi(\gamma) = \Psi(F_j(\gamma)) $  of a
Bargmann state  $ \Psi(\zeta_j) $  ``samples'' the values of  $\Psi$  only
on a sparse subset and we have to show that the  $ \Psi $  have a specific
functional form that enables one to reconstruct them uniquely from
their values at these ``sample points''.

We will proceed in two steps. First we show that there is no n-photon
state  $\Psi_n(\zeta_j)$  for which the image  $ \psi_n(\gamma) \equiv
{\cal T} \circ \Psi_n(\zeta_j)  $  vanishes on the loop space. In the
second step we will show that the same result holds for a general
Bargmann state.

Let us begin with a general 1-photon state, $ \Psi_1(\zeta_j) =
\int {d^3\k\over{\hbar|\k|}} \bar f_j(\k) \zeta_j(\k) $. We saw in the last
subsection that its image on the loop space is given by $\psi_1(\gamma)
= \oint_\gamma \bar A_a^{(f)} ds^a$ (see Eqs. (\ref{Aaf},\ref{Aaff})).
Now, if $\psi_1 = 0$ for all $\gamma$, the holonomy of
$ \bar A_a^{(f)} $ around any closed loop vanishes, whence $ A_a^{(f)} $
must be an exact 1-form.     However, from (\ref{Aaff}) it is clear that $
A_a^{(f)} $  is also divergence free, whence it must vanish identically. This
implies that $f_j(\k)$ must vanish, whence, $\Psi_1(\zeta_j)$ with which
we began must be zero. Thus, the restriction of the transform to a
1-photon state is indeed faithful. Let us next consider a general 2-photon
state, $ \Psi_2(\zeta_j) =
\int {d^3\k\over{\hbar|\k|}} \int {d^3\k'\over{\hbar|\k'|}} \bar f_{jj'}
(\k,\k') \zeta_j(\k)
\zeta_{j'}(\k') $. Its transform, $\psi_2(\gamma)$ is given by
$\psi_2(\gamma)=\oint_\gamma ds^a \oint_\gamma ds^{a'} \bar
A_{aa'}^{f_{2}} $, where  $ A_{aa'}^{f_{2}} (\vec
x, \vec x') $  is a ``2-point field'', satisfying
$ A_{aa'}^{(f_{2})} (\vec x, \vec x')  =  A_{a'a}^{(f_{2})}
(\vec x', \vec x) $,
given by the Fourier transform of  $ f_{jj'}(\k,\k') $. Let us suppose that
$ \psi_2(\gamma) = 0 $ for all holonomic  loops  $ \gamma $. Then,
by choosing for $\gamma$ loops $\alpha,\beta$  and  $\alpha\#\beta$, one
can readily show that
\f
        \oint_\alpha ds^a \oint_\beta ds^{a'}\ \bar A_{aa'}^{(f_{2})} = 0
\ff
for all loops  $\alpha,\beta$ . Now one can repeat the argument used
above for
1-photon states to conclude that $ f_{jj'}(\k,\k') $ and hence
$\Psi_2(\zeta_j)$  must vanish identically.
Thus, the transform is faithful also on 2-photon states. It is clear that
the argument can be extended, step by step, to n-photon states.

It  therefore remains to show that a linear combination,  $ \sum_n
\psi_n(\gamma) \equiv {\cal T} \circ \sum_n \Psi_n(\zeta_j)  $  , can
vanish only if  $ \sum_n \Psi_n(\zeta_j) $  itself vanishes. This is the
second step in the argument.   For simplicity, let us restrict ourselves to
helicity  $+1$  photons. The state   $ \Psi(\zeta_j)  $   is then of the
form
\f
        \Psi(\zeta_j) =
        \sum_{m=0}^n \int {d^3\k_1\over{\hbar|k_1|}}...\int {d^3\k_m\over
        {\hbar|k_m|}}
        \ f_n(\k_1,...,\k_m) \ \zeta_1(\k_1)... \zeta_1(\k_m)
\ff
whence its transform is of the form
\f
        \psi(\gamma) = \oint_\gamma ds_1 \dot\gamma^{a_2}(s_1) \ \ ...
\oint_\gamma ds_n \dot\gamma^{a_n}(s_n) \ \ \bar{A}_{a_1...a_n}^{(f_n)}
\ff
Let us suppose that  $\psi=0$  for all holonomic loop  $\gamma$. Choosing
for $\gamma$, loops $\alpha, \alpha\#\alpha, \alpha\#\alpha\#\alpha,
...$ , one obtains a series of constraints:
\begin{eqnarray}
        f_0 + \oint_\alpha ds^a \bar A_a +
        \oint_\alpha ds^{a_1} \oint_\alpha ds^{a_2} \bar A_{a_1a_2} +...+
        \oint_\alpha ds^{a_1} ... \oint_\alpha ds^{a_n} \bar A_{a_1...a_n}
        =0 ,
\nonumber \\
        f_0 + 2 \oint_\alpha ds^a \bar A_a +
        2^2 \oint_\alpha ds^{a_1} \oint_\alpha ds^{a_2}\bar A_{a_1a_2}+...+
        2^n \oint_\alpha ds^{a_1} ...\oint_\alpha ds^{a_n} \bar A_{a_1...a_n}
        =0,
\nonumber \\
        f_0 + 3 \oint_\alpha ds^a \bar A_a +
        3^2 \oint_\alpha ds^{a_1} \oint_\alpha ds^{a_2}\bar A_{a_1a_2}+...+
        3^n \oint_\alpha ds^{a_1} ...\oint_\alpha ds^{a_n} \bar A_{a_1...a_n}
        =0,
        \label{number}
\end{eqnarray}
etc. These infinite set of conditions can be satisfied if and only if
each term in the series vanishes, i.e.,
\f
 \oint_\alpha ds^{a_1} ...\oint_\alpha ds^{a_n} \bar A_{a_1...a_n} = 0
\ff
for all loop  $\alpha$  and all  $m\in[0,n]$ . Now, using the first step
in the argument we conclude that  $\Psi_m(\zeta_j)$  must vanish for all
$m$ , whence  $\Psi_m(\zeta_j)$  is itself zero. Thus, the transform is
faithful on the dense subspace of the Bargmann Hilbert space spanned by
all polynomials  $\Psi(\zeta_j)$. This is what we set out to prove. (Since
the domain space is dense, the transform can be extended to the full
Bargmann Hilbert space, preserving its faithful character. However, in
general the limit points are not expressible as functionals either in
the Bargmann or the loop pictures.)

It {\it is} somewhat surprising that although the set of the form factors
$F_j(\gamma,\k)$  is sparse  in the vector space of  $\zeta_j(\k)$, the
restriction   $\Psi(F_j(\gamma,\k)) \equiv \psi(\gamma)$  of
$\Psi(\zeta_j)$ determines  $\Psi(\zeta_j)$  completely.    How does this
come about? Two key features of loops, connections and Bargmann
representation  underlie this result. First, holonomic loops serve as a
``separating set'' for connections: if the holonomy of the two connections
around every loop is the same, the connections can differ at most by a
gauge transformation. This
feature is responsible for making the transform faithful on 1-photon
states. The second feature --which makes the transform faithful on the full
Fock space-- is an integral part of the Bargmann
representation: a general Bargmann state is a polynomial on the space
spanned by  $\zeta_j(\k)$ . It is because of this
property of $\Psi(\zeta_j)$ that none of them is annihilated by the
transform map  $ \cal T $. Perhaps a rough analogy would make this
point clearer. Consider the real line  $R$. Although integers form a sparse
set in $R$, the value of any polynomial
$f(x) = \sum_{m=1}^n a_n x^n$ on the set of the integers determines
$f(x)$  completely:  $f(m)=0$  for all integers  $m$  if and only if
$a_n=0$. The 1-photon Hilbert space, spanned by  $\zeta_j(\k)$  is
analogous to the real line in this example, and the polynomials, to
Bargmann states.

Since the map  $\cal T$  has no kernel, we can now use it to pull-back
Bargmann operators to operators on the loop space.   The images of the
creation and annihilation operators are:
\f
        {\cal T}\circ\hat \zeta_j(\k)\circ{\cal T}^{-1}\psi(\gamma) =
        F_j(\gamma,\k)\Psi(F_j(\gamma,\k)) =
        F_j(\gamma,\k) \psi(\gamma)
\ff
and
\f
        {\cal T}\circ\hat{\bar\zeta}_j(\k)\circ{\cal T}^{-1}\psi(\gamma) =
        \hbar|\k|\ {\delta\over\delta F_j(\gamma,\k)} \circ
        \Psi(F_j(\gamma,\k)) ,
\ff
where the functional derivative is well-defined because
$\Psi(F_j(\gamma,\k))$  is a polynomial in its argument. (In the next
section, we will be able to rewrite this term as a ``derivative''
operating directly on loop functionals  $\psi(\gamma)$ ).  It is now
straightforward to re-express  these operators,  such as the Hamiltonian,
the momentum and angular momentum operators on loop states.

There are two families of operators that are of particular interest to the
loop representation. The first is the holonomy  $\hat h[\gamma] =
\exp\oint_\gamma\m \hat{A}_a ds^a$  of the negative
frequency connection around the loop  $\gamma$ . From
(\ref{fouriercomp}), it follows that  $\m A_a\x$ can be expressed as
a sum of annihilation operators  $\hat{\bar\zeta}_j(\k)$ :
\f
        \m \hat A_a\x =
        {-1\over (2\pi)^{3/2}} \int {d^3\k\over{\hbar|\k|}}\
        e^{ i  \k \cdot \vec x}\
        \left[\hat{\bar\zeta}_1(-\k) m_a(\k)
        + \hat{\bar\zeta}_2(-\k) \bar m_a(\k) \right] ,
\ff
whence  $\hat h[\gamma]$  can be expressed as
\f
        \hat h[\gamma] = \exp \int {d^3\k\over \hbar|\k|}\
        F_j(\gamma,\k) \hat{\bar\zeta}_j(\k).
\ff
A straightforward calculation now yields the corresponding operator in
the loop representation. We have:
\f
        [{\cal T}\circ \hat h[\alpha]\circ {\cal T}^{-1} \psi](\beta) =
        \psi(\alpha\#\beta).
\ff
The action is thus surprisingly simple; it involves only the operation
$\#$,  which can be performed directly on the loop space, without, in the
end,  any reference to the transform or even the form factors of the loops.
It is  somewhat surprising that the operator is well defined because we
have  smeared the operator valued distribution  $\m \hat{A}_a\x$  only along
a  1-dimensional loop; we have effectively used a distribution --rather
than  a test field in  $C_0^\infty(\Sigma)$-- to smear  $\m \hat{A}_a\x$ .
Indeed,  had we used the real (or the positive frequency) connection in
the  definition of  $\hat h[\alpha]$, the resulting operator would {\it not}
have  been well-defined,  it would have mapped any n-photon state to a
``state  with infinite norm''. This somewhat unexpected ``tame'' behavior
of the operator in the Abelian theory may be
related to the surprisingly simple renormalization properties of the
Wilson-loop of the bare (non-renormalized) field operators in Yang-Mills
theories discussed, e.g., in \cite{14bis}.

Since the holonomy operators $\hat h[\alpha]$ will play an important
role in the direct construction of the loop representation in the next
section, let us examine their eigenvalues and eigenvectors. Since they are,
in essence, exponential of annihilation operators, one would expect their
eigenvectors to be coherent states. This is indeed the case. Fix a
transverse connection  $C_a\x$  for which  $\int {d^3\k\over{\hbar|\k|}}
|C_a(\k)|^2 <\infty$  and consider the loop state
\f
        \psi_C(\gamma)=\exp \oint_\gamma C_a ds^a.
\ff
Then we have:
\f
        [{\cal T}\circ \hat h[\alpha]\circ {\cal T}^{-1} \psi_C]\ (\beta) =
        \left( \exp \oint_\alpha C_a ds^a \right)  \psi_C(\beta).
\ff
Thus, $\psi_C(\beta)$ is a simultaneous eigenstate of {\it every} holonomy
operator. Therefore it is the coherent state peaked at the connection $C_a\x$.
Note, incidently, that the states $|\gamma\rangle$ --the ``basis states'' of
the loop representation-- introduced at the end of Sec 3.2 are specific
examples of these coherent states: They are the coherent states corresponding
to (real, transverse) connections which are concentrated along the loops.

The second interesting family of operators is constructed from the
positive frequency electric field $\p \hat E^a\x$.  Set
\f
        \hat E[f] \equiv \int \p \hat E^a\x \bar f_a\x,
\ff
where  $f_a$  is an equivalence class of $C^\infty$ 1-form of compact
support, modulo exact 1-forms. From (\ref{fouriercomp2}) it follows that
\f
        \hat E[f] =
        {-i\over (2\pi)^{3/2}} \int {d^3\k\over{\hbar|\k|}}\
        \bar f_j(\k) \hat\zeta_j(\k),
\ff
where we have set  $f_a\x = {1\over (2\pi)^{3/2}} \int {d^3\k
\over{\hbar|\k|}}\ \ (f_1(\k) m_a(\k)+f_2(\k) \bar m_a(\k)+f_L(\k) k_a )$.
Using the definition (\ref{abtransf}) of the transform, it now follows that
the corresponding operator in the loop representation is simply
\f
        [{\cal T}\circ \hat h[\alpha]\circ {\cal T}^{-1} \psi](\gamma) =
        i\hbar\ \left( \oint_\gamma f_a ds^a \right)  \psi(\gamma).
\ff
Again, the action of the operator on loop states is rather simple, one does
not need form factors to express it. (That the operator is well-defined is
not surprising because  $\p \hat E^a\x$  has been  smeared by a
$C^\infty$  test field of compact support). By inspection, the loop states
\f
        \psi_{\gamma_0}(\gamma) =
        \left\{
        \begin{array}{ll}
                1 & \mbox{if}\ \gamma=\gamma_0 \\
                0 & \mbox{otherwise}
        \end{array}
        \right.,
\ff
which are characteristic functions of a given loop  $\gamma_0$  are
simultaneous (generalized) eigenvectors of $\hat E[f]$ for all  $f_a$ :
\f
        \hat E[f]  \circ \psi_{\gamma_0}(\gamma) =
       i\hbar \left(\oint_{\gamma_0} f_a ds^a  \right) \
        \psi_{\gamma_0}(\gamma) .
\ff
These states do not belong to the Fock space since they are not
normalizable. This is not surprising because  $\hat E[f]$  is essentially
a superposition of creation operators. Note that these states,
$\psi_{\gamma_0}(\gamma) $, are ``complementary" to the coherent states
$\psi_C(\gamma) $ --and hence, in particular, to the states $|\gamma\rangle$
introduced in section 3.2-- which are eigenvectors of $\hat h[\alpha]$, i.e.
of superpositions of annihilation operators. Indeed, while $\psi_{\gamma_0}
(\gamma) $  have support on a single loop  $\gamma_0$, the states
$\psi_C(\gamma) $ are spread out in the loop space. Similarly, while
$\psi_C$ are concentrated at a single connection $C_a$, the states
$\psi_{\gamma_0}$ are completely spread out in the space of
connections. This suggest that the two sets of operators, $\hat h[\alpha]$
and  $\hat E[f] $, should be conjugate to one another in a suitable sense.
This is indeed the case: they obey the commutation relations:
\f
        [\hat h[\alpha],\ \hat E[f] ] =
        \left(i\hbar\ \oint_\gamma f_a ds^a \right)\ \ \hat h[\alpha].
\ff
The fact that the commutator is closed and has such a simple expression
suggests that the algebra of these operators may serve as a useful point
of departure for the construction of the loop representation. We will see
in the next section that this expectation is indeed correct.

\section{Loop representation}

In this section we construct a loop representation of the Maxwell field
directly, without any reference to the Bargmann representation of section
2 or the loop transform of section 3. It is this construction that appears
to best suited for use in lattice QCD \cite{6} and canonical quantum
gravity \cite{2}. The final picture we arrive at is the same as that
obtained in section 3. Thus the representation we now construct from
first principles is isomorphic with the Bargmann --and hence, also Fock--
representation.

This section is also divided in to three parts. In the first we sketch
the general quantization program \cite{15,16} that we will follow, a program
that is also applicable to Yang-Mills' and Einstein's theories. In the second
we construct the algebra of quantum operators, the loop algebra.  In the
third we construct a representation of this algebra on a suitably defined
space of functionals of holonomic loops and and discuss some features of
the resulting loop representation.

\subsection{Quantization program}

Consider a classical system with phase space $\Gamma$. To quantize the
system we proceed in the following steps: \footnote{For a more complete
discussion of the program including a treatment of constrained systems,
see Refs. 15 and 16.}

1. Choose a subspace $\cal S$ of the space of complex valued functions on
$\Gamma$ which is closed under the Poisson bracket operation and large
enough so that any well-behaved function on $\Gamma$ can be
expressed as (possibly the limit of) a sum of products of elements of
$\cal S$. Elements of $\cal S$ are referred to as {\it elementary classical
variables} and are to have unambiguous quantum analogs. In the
Bargmann quantization, for example, $\cal S$ is the vector space spanned
by functionals $\int {d^3\k\over{\hbar |\k|}} \bar f_j(\k) \zeta_j(\k),
\ \ \int {d^3\k\over{\hbar |\k|}} g_j(\k) \bar\zeta_j(\k)\ $, which are
linear in $\zeta_j(\k)$ and $\bar\zeta_j(\k)$ respectively, and by constants.

2. Associate with each $f$ in $\cal S$ an {\it elementary quantum
operator} $\hat f$ and consider the free associative algebra generated by
these abstract operators. Impose on this algebra the (generalized)
canonical commutation relations:
\f
          [\hat f,\ \hat g] =  i \hbar \ \widehat{\{f,g\}}
\ff
for all $f$ and $g$ in $\cal S$. In addition, if the set $\cal S$ is
overcomplete, impose on the algebra also ``anti-commutation  relations'',
namely the relations that capture the algebraic relations that exist
between elements of $\cal S$.  For instance if $f$, $g$ and $h=fg$ are
all in $\cal S$, then
\f
        \hat f \cdot \hat g + \hat g \cdot \hat f = 2 \ \hat h.
\ff
Denote the resulting associative algebra by $\cal A$. (Note that at this
stage $\cal A$ is an abstract algebra: there is, as yet, no Hilbert space
for it to act upon.)

In the example of the Bargmann representation, the canonical
commutation relation reduce to:
\f
        \left[
        \int {d^3\k\over |\k|}\  \bar{f}_j(\k)\hat\zeta_j(\k),\
        \int {d^3\k\over |\k|}\  g_j(\k) \hat{\bar\zeta_j}(\k) \}
        \right]
        = \hbar \int {d^3\k\over |\k|}\ \bar{f}_j(\k)g_j(\k) .
\ff
There are no anti-commutation relations because the set $\cal S$ is
complete but not overcomplete: there are no algebraic relations among
the elements of $\cal S$. The loop variables we want to introduce in this
section, on the other hand, are overcomplete and anti-commutation
relations will be important.

3. Introduce an involution, $*$, on $\cal A$ by setting
\f
        (\hat f)^* = \hat{\bar f\ }
\ff
for all elementary variables $f$ (the bar denotes complex conjugation
as before) and requiring that $*$ satisfies the defining properties of
an involution: $(\hat A+ \lambda \hat B)^* = \hat A + \bar\lambda\hat B^*;
\ \ (\hat A\hat B)^*  = \hat B^* \hat A^*$  and $(\hat A^*)^* = \hat A$,
for all $\hat A, \hat B$  in $\cal A$ and complex numbers $\lambda$.
Denote the resulting $*$--algebra by $\cal A^{(*)}$.

4. Choose a linear representation of $\cal A$ on a complex vector space
$V$. (The $*$--relations are ignored at this step.) In the Bargmann case,
$V$  is the space of polynomials $\Psi(\zeta_j(\k))$, the operators
$\int {d^3\k\over |\k|}\  \bar f_j(\k) \hat\zeta_j(\k)$ are represented by
the multiplication operators and $\int {d^3\k\over |\k|}\  g_j(\k)
\hat{\bar \zeta}_j(\k)$ by the functional derivative along $g_j(\k)$.

5. Introduce on $V$ an inner product $<\ ,\ >$ so that the quantum reality
conditions are satisfied:
\f
        <\Psi,\ \hat A \Phi>\  =\  <\hat A^*\Psi,\ \Phi>
\ff
for all $\Psi$ and $\Phi$ in $V$ and $\hat A$ in $\cal A^{(*)}$. Thus,
it is the $*$--relations that are to select the inner product. In section 2
we saw that the strategy does indeed pick out an unique inner product on
the Bargmann states.

The program requires two external inputs: the choice of $\cal S$ in step
1 and the choice of of the carrier space $V$ of the representation in step
4. One  may make ``wrong'' choices and find that the program cannot be
completed.  However, if the choices are viable, i.e., if the program can be
completed at all, the resulting inner product is unique on each irreducible
sector of the representation of ${\cal A}$ on $V$.
In the framework of this program, the textbook treatments of field
theories correspond to choosing for elements of $\cal A$ the smeared-out
field operators, and, for $V$, the Fock space or, alternatively,
suitable functionals of fields. In the loop quantization, on the other hand,
one changes this strategy.\cite{2} \ \ Both $\cal S$ and $V$ are now
constructed from holonomic loops.

\subsection{A loop algebra}

In this sub-section, we will carry out the first three steps in the above
quantization program.

The discussion at the end of section 3 suggests that we choose as our
elementary classical variables the complex-valued functionals $h[\alpha]$
and $E[f]$, labelled by holonomic loops $\alpha$ and equivalence classes
of 1-form $f_a$ (where $f_a\approx f_a+\partial_a g$ for any $g$) on a
spacelike 3-plane $\Sigma$, defined via
\f
        h[\alpha] := \exp \ \oint_\alpha \m A_a ds^a
        \label{ha}
\ff
and
\f
        E[f] := \int_\Sigma \p E^a \, f_a \ d^3\vec x .
        \label{Ef}
\ff
Here, $\m A_a$ is the negative frequency part of the real, transverse
connection $A^T_a$ and $\p  E^a $ is the positive frequency part of
the real, transverse electric field $E_T^a$. Thus,
$      \m A_a\x  = {1\over\sqrt{2}}
      \left(A^T_a\x -  i \, \Delta^{- 1/2} E_{T\,a}\x\right)  $
and $   \p E^a\x  = {1\over\sqrt{2}}
        \left(E_T^a\x - \right. $
     $\left. i  \Delta^{1/2} A^{T\,a}\x\right)  $
(Eqs. (\ref{fouriercomp},\ref{fouriercomp2}) \ ).
The knowledge of the holonomies $h[\alpha]$ for all $\alpha$ enables one
to reconstruct $\m A_a\x$ completely, while the knowledge of $E[f]$
for all $f_a$ suffices to determine $\p  E^a\x$.  Hence,  $h[\alpha]$
and $E[f]$, provide us with an (over-) complete coordinatization of the
phase space. The space $\cal S$ of elementary classical variables required
in the first step of the quantization program shall be the vector space
generated by the  $h[\alpha]$ and $E[f]$, subject to the obvious linear
relations
\f
        E[f_1] + E[f_2] = E[f_1+f_2],
\ff
for all $f_1$ and $f_2$.
It is closed under the Poisson-bracket operation because:
\f
        \{h[\alpha],\ E[f]\} = \left(\oint_\alpha f_a ds^a \right) h[\alpha].
\ff
For future use, we note that the negative frequency magnetic fields $\m
B^a \x \equiv \epsilon^{abc}\partial_b \m  A_a \x $ can
be recovered from the holonomies $h[\alpha]$ by the obvious limiting
procedure.

The next step in the quantization program is the construction of the
algebra  $\cal A$ of quantum operators.  Let us associate with each
$h[\alpha]$ in  $\cal S$ an operator $\hat h[\alpha]$ and with each $E[f]$
an operator $\hat E[f]$ and consider the associative algebra generated by
finite sums of products of these elementary quantum operators. As
mentioned in section 4.1, this is an abstract algebra; there is as yet no
vector space for it to  act upon. On this algebra, impose the canonical
commutation relations:
\f
        [\hat h[\alpha], \hat h[\beta] ]  =  0 ,  \ \ \
        [ \hat E[f], \hat E(g)]  =  0 ,
        \nonumber
        \label{ccr}
\ff \f
        [ \hat h[\alpha], \hat E(g) ]  =
         i \hbar\ \left(\oint_\alpha f_a ds^a \right)\ \hat h[\alpha].
        \label{CCR}
\ff
Further, we must incorporate in this quantum algebra the fact that
$h[\alpha]$ is over-complete, i.e. that there are algebraic relations
among them: $h[\alpha] h[\beta] = h[\alpha\#\beta]$. This is achieved by
imposing on the algebra the relations
\f
        \hat h[\alpha] \hat h[\beta] = \hat h[\alpha\#\beta]
        \label{agebraicrel}
\ff
for all holonomic loops $\alpha$ and $\beta$. (We do not need
anti-commutators because the $\hat h[\alpha]$ commute.) The result is
the algebra $\cal A$ of quantum operators.\footnote{More
precisely, we consider the two sided ideals generated by equations
(\ref{ccr},\ref{CCR},\ref{agebraicrel}) and take the quotient of the
associative algebra by it.}

In analogy with the terminology
used in gravity \cite{2,8}, we will call $\cal A$ a {\it loop algebra}.
Finally, let us impose the $*$-relations. Since the transverse fields
$A^T_a\x$ and $E_T^a\x$ in the reduced phase space are all real, it
follows immediately that the complex conjugate of $\p  E^a\x $ is
determined by the negative frequency magnetic field
$\m B^a\x$  via
\f
        \overline{\p E^a\x}=
        i \Delta^{-1/2}\mbox{curl}\,\m B^a\x.
\ff
Hence, in the quantum theory, the $*$-relations are given by
\f
        \left( \hat E[f] \right)^* =
        i  \hat B[\mbox{curl}\,\Delta^{- 1/2}\bar f],
        \label{starrelations}
\ff
where $\m \hat B^a(f)$ is to be obtained by taking the obvious limit
of the holonomy operators $\hat h[\alpha]$ and where, on the
right side, we have used the unique transverse 1-form in the equivalence
class to which $f_a$ belongs.  In the next section, we introduce a linear
representation of $\cal A$, find an inner product such that the
$*$--relations (\ref{starrelations}) are realized by the concrete operators
representing electric fields $\hat E[f]$ and holonomies $\hat h[\alpha]$.

\subsection{Loop states}

We can now represent the quantum algebra $\cal A$ by operators acting
on suitably regular functionals $\psi(\gamma)$ on the holonomic loop
space $\cal HL$.  The discussion at the end of section 3.3 will motivate
the various steps in the construction.

The idea is to construct the loop states simply from sums of products of
$\psi(\gamma)$ of the type $\psi(\gamma) = \oint_\gamma f_a ds^a$.
For a precise implementation of this idea, we proceed as follows.  Denote
by $\cal V$ the vector space of equivalence classes of complex valued,
smooth 1-forms $f_a\x$ of compact support, where $f_a\approx
f_a+\partial_a g$ for any smooth function $g$ of compact support.
Denote by ${\cal V}^{n}$  the totally symmetric, n-th rank tensor
product of $\cal V$ with itself, spanned by the fields
$f_{a_1...a_n}(\vec x_1,...,\vec x_n)$.  (The symmetry property requires
that  $f_{a_1a_2}(\vec x_1,\vec x_2) = f_{a_2a_1}(\vec x_2,\vec x_1) $,
and so on.) The carrier space $V$ underlying our representation of the
loop algebra  $\cal A$ is then spanned by the functionals $\psi(\gamma)$
on $\cal HL$  of the type
\begin{eqnarray}
        \psi(\gamma) =   & & \bar f_0
\nonumber \\ & &
        +\oint_\gamma ds^a \ \bar f_a
\nonumber \\ & &
        +\oint_\gamma ds^{a_1}\oint_\gamma ds^{a_2}\ \bar f_{a_1a_2}
\nonumber \\ & &
        +\ \  ...
\nonumber \\ & &
        +\oint_\gamma ds^{a_1}...\oint_\gamma ds^{a_n}\ \bar f_{a_1...a_n}.
        \label{states}
\end{eqnarray}
Here $n$ is an arbitrary (but finite) integer, $f_0$ is a complex number,
$f_{a_1...a_n}$ belongs to ${\cal V}^n$, and we have used the
same notation as in (\ref{notation}) for integrals.
Note that, although one may use parametrization of the loop
$\gamma$ to write out the functional form of $\psi(\gamma)$ explicitly,
the resulting functional is independent of the parametrization;
$\psi(\gamma)$ depends only on the holonomic loop $\gamma$.
Similarly, the value of $\psi(\gamma)$ remains unaltered if any of the
$\bar f_{a_1...a_n}$ is replaced by any of its gauge equivalent fields,
e.g. if $\bar f_{a_1,a_2}(\vec x_1, \vec x_2)$ is replaced by
$ \bar f_{a_1,a_2}(\vec x_1,\vec x_2) +  \bar{f}_{a_1}
(\vec x_1)\partial_{a_2}g(\vec x_2) + \bar{f}_{a_2}(\vec x_2)\partial_{a_1}
g(\vec x_1)$. Finally recall that any holonomic loop is completely
determined by its form factors $F_j(\gamma,\k),\ \ j=1,2$. We can therefore
express $\psi(\gamma)$ of equation (\ref{states}) as a functional of these
form factors. It is simply a {\it polynomial\ } in $F_j(\gamma,\k)$:
\[
        \psi(\gamma) =
\] \f
        \sum_{m=1}^n
        \int {d^3\k_1\over{\hbar |k_1|}}...\int {d^3\k_m\over{\hbar |k_m|}}
        \bar f_{j_1...j_m}(\k_1,...,\k_m)
        F_{j_1}(\gamma,\k_1) ... F_{j_m}(\gamma,\k_m) ,
\ff
where $f_{j_1...j_m}(\k_1,...,\k_m)$ are the Fourier coefficients of
the  transverse (in all indices) parts of $f_{a_1...a_1}(\vec x_1,...,\vec
x_n)$.  This provides an equivalent characterization of the carrier space
$V$.

Next we must represent $\cal A$ by concrete operators on $V$.  The
results of section 3 suggest the following two definitions:
\begin{eqnarray}
        \left(\hat h[\alpha]\circ\psi\right)(\beta)
        & := & \psi(\alpha\#\beta)
        \label{h}
\\
        \left(\hat E[f]\circ\psi\right)(\beta)
        & := & i\hbar\, \left(\oint_\beta f_a ds^a\right)\, \psi(\beta)
        \label{E}
\end{eqnarray}
(for simplicity of notation we will not distinguish between elements of
$\cal A$ and operators on $V$ representing them). It is straightforward
to verify that (\ref{h}) and (\ref{E}) are well-defined linear mappings
on $V$ and that they provide a representation of the generalized CCR's
(\ref{CCR}) and algebraic relations (\ref{agebraicrel}). Since elements of
$\cal A$ are sums of products of $\hat h[\alpha]$ and $\hat E[f]$ subject
to these relations, we have constructed a representation of the loop
algebra $\cal A$ by linear operators acting on suitable functionals on the
holonomic loop space. This completes the fourth step of the quantization
program.

In the fifth and the last step we must use the $*$--relations
(\ref{starrelations}) to select an inner product on $V$; since the
$*$--relations are expressed in terms of the magnetic field operator
${}^{(-)}\!\hat B^a\x$, we must first construct this operator on $V$.
Let us begin with classical holonomies. Given a point $\vec x $ and a
unit vector $t^a$ at $\vec x$, the negative frequency magnetic field
$\m B^a\x $ can be recovered from holonomies via:
\f
        t_a\, \m B^a\x = \lim_{\epsilon\to 0} {1\over\epsilon}
       (h[\alpha(t,\epsilon,\vec x)]- 1)
\ff
where $ \alpha(t,\epsilon, \vec x)$ is a loop of area $\epsilon$ in
the 2-plane perpendicular to $t^a$, centered in $\vec x$.
The action of the operator valued distribution $\m \hat B^a\x$ on
$V$ is therefore defined by
\f
        \left(t_a\, \m \hat B^a\x\right)\psi(\gamma) =
        \lim_{\epsilon\to 0} {1\over\epsilon}
        \left[
        \psi(\gamma\#\alpha(\epsilon,t,\vec x))- \psi(\gamma)
        \right].
        \label{B}
\ff
(Note that $ \m \hat B^a\x $ is a derivative operator on loop space; it
is not  a functional derivative, nor it is the area derivative loop
operator \cite{loops}.
Rather, it is essentially the generator of the loop group
introduced by Gambini and Trias \cite{abeliangambini}.)
Using this expression, it is now straightforward to compute the
commutator of $\hat B[g] := \int \m \hat B^a\x g_a\x d^3\vec x$ and
$\hat E[f]$:
\f
        [\hat B[g],\ \hat E[f]] =
        i\hbar\left(\int (\mbox{curl}\,g^a)f_a d^3\vec x\right)\
        1\!{\rm I}.
        \label{commBE}
\ff

We can now determine the inner-product. Consider the state $\psi_0$
defined by $\psi_0(\gamma )= 1$ and fix normalization by setting
$<\psi_0,\,\psi_0> = 1$. This state is clearly cyclic in $V$. More
precisely, since
\f
        (\hat h[\alpha]\circ\psi_0)(\gamma) = \psi_0(\gamma)
\label{honvac}
\ff
and
\f
        (\hat E[f]\circ\psi_0)(\gamma) = i\hbar
        \left(\oint_\gamma f_a ds^a\right),
\ff
it is clear that any element in $V$ can be obtained by taking linear
combinations of states obtained by acting on $\psi_o(\gamma)$
repeatedly by the electric field operator $\hat E[f]$. Thus, the
representation mapping given by (\ref{h}) and (\ref{E}) naturally leads
us to regard $\psi_0$ as the vacuum state, $\hat h[\alpha]$ as the
exponential of the annihilation operators and $\hat E[f]$ as the
creation operators. The discussion of section 3 makes it clear that this
interpretation is correct. However, we did not have to refer to the loop
transform to arrive at this picture; the loop representation introduced
here leads us directly to these conclusions.

Let us use the quantum reality conditions to determine the inner product
between first excited states, $\hat E[f]\psi_0$ and $\hat E[g]\psi_0$.
Using the $*$-relations (\ref{starrelations}) and the commutation
relations   (\ref{commBE}), we have
\begin{eqnarray}
        <\hat E[f]\psi_0,\ \hat E[g]\psi_0> & = &
        <\psi_0,\ i\hat B(\mbox{curl}\Delta^{-1/2}\bar f) \hat E(g) \psi_0>
\nonumber \\
        & = & <\psi_0,\
        [i\hat B(\mbox{curl}\Delta^{-1/2}\bar f), \hat E(g)] \psi_0>
\nonumber \\
        & = &\hbar \int (\bar f^a \Delta^{1/2}g_a) d^3\vec x,
\end{eqnarray}
where, we have used the fact that $\hat{B}[f ]$ annihilates $\psi_0$
(see Eq.(\ref{honvac})) and, as before, carried out the calculation in
the transverse gauge. Let us define $f_a(\k)$ via
\f
        f_a\x = {1\over{(2\pi)^{3/2}}}
        \int {d^3\k\over{\hbar |\k|}}\ f_a(\k)\ e^{i\k\cdot\vec x}.
        \label{ft}
\ff
Then, we can write the inner-products between the first excited states
simply as:
\begin{eqnarray}
        <\psi_f(\gamma),\ \psi_g(\gamma)> & \equiv &
        <\oint_\gamma f_a ds^a,\ \oint_\gamma g_a ds^a>
\nonumber \\
        & = & \hbar \int {d^3\k\over{\hbar|\k|}}\ \bar f_a(\k) g^a(\k).
\end{eqnarray}
It is straightforward to repeat this calculation for higher excited states.
The norm of the general element $\psi(\gamma)$ defined in (\ref{states}) is
given simply by
\begin{eqnarray}
        ||\psi(\gamma)||^2 & = &
\nonumber \\
        & & |f_0|^2 +\hbar \int {d^3\k\over{\hbar |\k|}}\ |f_a(\k)|^2
\nonumber \\ & &
        +\hbar^2 \int {d^3\k_1\over{\hbar |\k_1|}} \int {d^3\k_2\over
        {\hbar |\k_2|}}\
        |f_{a_1a_2}(\k_1,\k_2)|^2 + ...
\nonumber \\ & &
        +\hbar^n \int {d^3\k_1\over{\hbar |\k_1|}} ... \int {d^3\k_n\over
        {\hbar |\k_n|}}\ \ |f_{a_1...a_n}(\k_1,...,\k_n)|^2
\label{norm}
\end{eqnarray}
where $f_{a_1...a_m}(\k_1,...,\k_m)$ is the Fourier transform of the
transverse part of $f_{a_1...a_m}(\vec x_1,...,\vec x_m)$ defined via
(\ref{ft}). Thus the quantum reality conditions do indeed suffice to
determine the inner-product on $V$. The Hilbert space of states is the
Cauchy completion $\bar V$ of this $V$. It is clear from section 3.3 that
$\bar V$ is isomorphic with the standard Fock space of photons. Again,
we were able to single out the correct scalar product without any direct
reference to Poincar\'e group.

{}From the viewpoint of rigor, however, this construction is unsatisfactory
in one respect: to write out the $*$-relations on the algebra ${\cal A}$,
we had to refer to the magnetic field which belongs {\it not} to the algebra
${\cal A}$ but to a suitable (unspecified) completion thereof. However,
this incompleteness refers to the {\it constructive procedure} used in
this sub-section rather than to the arguments that establish the {\it
existence} of the loop representation and its equivalence to the Bargmann
(or, Fock) picture. More precisely, we can just work with the algebra
${\cal A}$ introduced in section 4.2, refrain from introducing the
$*$-relations, and simply exhibit a representation of ${\cal A}$: states
are given by (\ref{states}), the inner-product by (\ref{norm}) and the
generators of ${\cal A}$ are represented via (\ref{h}) and (\ref{E}). This
gives the complete quantum description. In particular, the $*$-relations
could now be deduced {\it from} the inner-product. However, now the inner
product itself has to be simply postulated. In this sense, the new
procedure would fail to be constructive. In the spirit of the general
quantization program \cite{15}, one would like to ``derive'' the correct
inner product by using the quantum reality conditions and to implement
these conditions one must introduce on ${\cal A}$ a $*$-relation. Therefore,
to complete the discussion presented in this section, we must either work
with an appropriate completion of the algebra ${\cal A}$ which includes
the magnetic field operator, or, alternatively, replace the algebra by
another one which {\it is} closed under the $*$-relations without the need
of any completion. The second of these strategies has been completed
successfully. (The new algebra is based on ``thickened-out loops''.) It will,
however, be presented elsewhere since the discussion involved is somewhat
long.

To conclude this section, let us exhibit the Hamiltonian operator directly
in the loop representation. The classical Hamiltonian can be expressed in
terms of $\p E(\vec x )$ and its complex conjugate as:
\f
        H\ = \int_\Sigma d^3{\vec x}\ \ \p E^a \x\ \ \overline{{}^+\!E_a}\x .
\ff
Since $\hat E$ is a creation operator, the normal ordered quantum
Hamiltonian can be written as
\f
       \hat H \ = \int_\Sigma d^3{\vec x}\ \hat E^a\x \ (\hat E_a\x)^*.
\ff
Using the reality condition (\ref{starrelations}) we can substitute for
$(\hat{E}_a)^*$ in terms of the magnetic field operator $\hat{B}_a$. The
momentum space expression of the Hamiltonian operator is then given by
\f
       \hat H \ = \int {d^3\k\over {|\k|}}\ \epsilon_{abc}\ \hat{E}^a(-\k)
       \ \hat{B}^b(\k)\  k^c .
\ff
To find its eigenvectors and eigenvalues, we first note the action of the
magentic field operator $\hat{B}^a(\k )$ on 1-photon states, $\psi (\gamma )
= F^b(\gamma, -\k')$, with momentum $-\k'$:
\f
    \hat B^a(\k)\cdot F^b(\gamma,-\k') =
        i\epsilon^{abc} k_c \delta^3(\k,\k'),
\ff
which follows directly from the definition (\ref{B}) of $\hat B^a$. Using
this result in the expression of the Hamiltonian, we have:
\f
   \hat{H}\cdot F^b(\gamma, -\k') = \hbar |{\vec k}'| F^b(\gamma, -\k').
\ff
Thus, as expected, these photon states are eigenstates of the Hamiltonian
with eigenvalue $\hbar|{\vec k}|$. More generally, the $n$-photon states,
$\psi(\gamma )= F_{j_1}(\gamma, \k_1)....F_{j_n}(\gamma, k_n)$,
with momenta $\k_1,...\k_n$ and polarizations $j_1...j_n$ are eigenstates
of $\hat{H}$ with eigenvalue $\hbar \sum_i |\k_i|$.

Finally, it is interesting to note the curious form that the number
operator $\hat N$ takes in the loop representation. From equation
(\ref{number}) it is easy to see that $\hat N$ is given by
\f
        2^{\hat N} \psi(\alpha) = \psi(\alpha\#\alpha).
\ff

\section{Discussion}

In this paper, we first recast the Bargmann description of free photons in
terms of loops using a functional transform and then showed that the
resulting loop representation can be obtained directly in the framework
of a general quantization program. It is this later strategy that was first
used in non-perturbative gravity \cite{2}. The fact that one can recover
the Fock description of photons working entirely with holonomic loops
and 1-forms provides new support to the methods used in that program.
It is also clear from the discussion of sections 3 and 4
that connections and loops provide equivalent ways of handling gauge
theories. The emphasis is, however, different: while the holonomy
operators are diagonal in the connection representation, it is the electric
field operators that act by multiplication in the loop representation.

It is important to note that that there is not one but {\it many} loop
representations. To see this, consider first non-relativistic quantum
mechanics. There, one may introduce on the Hilbert space of
square-integrable functions $\Psi(\lambda)$, introduce on it the operators
$\hat{A}$ and $\hat{B}$ via $\hat{A}\cdot \Psi(\lambda )= \lambda
\Psi(\lambda )$ and $\hat{B}\cdot\Psi(\lambda) = -i\hbar d\Psi(\lambda )
/d\lambda$, and verify that they satisfy the Heisenberg commutation
relations. This information does not, however, suffice to determine the
physical interpretation of the states $\Psi(\lambda )$ and the operators
$\hat {A}, \hat{B}$. For example, if $\lambda$ is identified with $x$,
$\hat{A}$ has the interpretation of the position operator and $\hat{B}$,
of the momentum operator. In this case, the states $\Psi_\alpha (\lambda)
=\ \exp  \ i\alpha\lambda$ are the eigenstates of the momentum operator
with eigenvalue $\hbar\alpha$. If, on the other hand, $\lambda$ is identified
with $p$, it is $\hat A$ that represents the momentum operator and $-\hat{B}$
that represents the position. The states $\Psi_\alpha (\lambda)$ are now
eigenstates of the position operator with eigenvalue $-\hbar\alpha$. The
situation is similar with the loop presentation. In the Abelian case
considered in this paper, one can begin with functions $\Psi(\gamma)$ of
holonomic loops $\gamma$, define on this space two operators $\hat{h}
[\alpha]$ and $\hat{E} [f]$ via Eqs. (\ref{h}) and (\ref{E}) and verify
that they satisfy the (generalized) CCR. As in the case of non-relativisitic
quantum mechanics, by itself this procedure does not determine the
representation uniquely. We have to supply, in addition, the interpretation
of these operators. In this paper, we let $\hat{h}[\alpha ]$ be the holonomy
operator defined by the {\it negative frequency} connection and $\hat{E}[f]$
be the smeared out {\it positive frequency} electric fields. One could have
made other choices and thus obtained other loop representations. For example,
we could have let $\hat{h}[\alpha]$ be the holonomy of the {\it real}
connection and $\hat{E}[f]$ be the smeared out {\it real} electric fields.
The resulting description would then have been equivalent to that obtained
by Gambini and Trias \cite{abeliangambini}. Alternatively, we could have
let $\hat{h}[\alpha ]$ be the holonomy of the self dual connection and
$\hat{E}[f]$ be the anti-self dual electric field. This avenue turns out
to be the most useful one in canonical gravity. (For treatment of linearized
gravity --i.e. gravitons-- along these lines, see \cite{8}.) The general
choice is to let $\hat{h}[\alpha ]$ be the holonomy of a certain kind of
connection and $\hat{E}[f]$, the electric field canonically conjugate to
that connection. The choice we made in this paper is the simplest one:
with this choice, the transform of section 3 as well as the action of the
operators in the resulting loop representation are well-defined without any
need of regularization. The situation is more complicated in other loop
representations. In particular, as we pointed out earlier, since the holonomy
operator is smeared only in one dimension --along the loop-- it fails to
be well-defined on the Fock space if the connection is real or self dual and
regularization involving ``thickened-out loops'' then becomes {\it essential}.
In all cases, at least formally, the loop states $\psi(\gamma )$ can be
thought of as the matrix elements $<0|\hat{h}[\gamma]|\Psi>$ of the holonomy
operators between the vacuum state and a given state $\Psi$. It is only in
the representation analysed in this paper that $\hat{h}[\gamma]$ are
the exponentials of annihilation operators and the interpretation is precise.
Furthermore, now $\psi(\gamma )$ can, in addition, be interpreted in terms
of a coherent state basis.

An interesting feature of any loop representation is the way in which it
deals with gauge invariance. In section 4.2, we defined $h[\alpha]$ and
$E[f]$ in terms of the {\it transverse} parts of the positive and negative
frequency fields (Eqs. (\ref{ha}) and (\ref{Ef}). Note, however, that we
could have dropped the restriction to the transverse part without changing
the final result. For, the holonomy of a connection depends only on its
transverse part and, by its very definition (\ref{E}), the quantum operator
$\hat E[f]$ also depends only on the transverse part of $f$. In the
non-abelian case, the details of the situation are more complicated. However,
it is again true that gauge invariance is {\it automatically} implemented
in the loop representation and does not have to be imposed as a restriction
on quantum states. Thus, the loop representation captures the physical
degrees of freedom of a gauge theory in a natural way. It is this simplicity
that provided a primary motivation for the use of loop space methods in
gauge theories by Madelstam, Migdal, Polyalov, Wilson and others
\cite{loops} over the past three decades.

In a sense, however, the tradition of using loops as basic objects goes
back substantially further --in fact, all the way to Faraday! For, gauge
theories can be said to have originated in Maxwell's work which formalized
Faraday's intuitive picture of electromagnetism as a theory of ``lines of
force'' trapped in space. In absence of sources, each line of force is a
closed loop. It turns out, quite remarkably, that this picture of a
classical field has direct analogs in the loop formulation of the quantum
theory.

To see this, recall first that there exist two quantum states that can be
naturally associated with any given loop $\gamma$. First is the (generalized)
coherent state, $ C_{F(\gamma)}(\zeta_j) \equiv \exp\int {d^3\k\over{\hbar
|\k|}} \bar
F_j(\gamma ,\k)  \zeta_j(\k) $. We saw at the end of section 3.2 that the
expression $\Psi(\gamma )$ of a state in the loop representation can be
interpreted as the scalar product between the state $\Psi$ and
$C_{F_j}(\zeta_j)$. Now, this coherent state is peaked precisely at the
classical field configuration in which the (real, transverse) connection is
concentrated along the loop $\gamma$. Thus, the number $\Psi(\gamma_o)$
represents the probability amplitude of seeing a loop-like classical
excitation of the connection along $\gamma_o$ when the quantum system is
in the state $\Psi$. The second state associated with a closed loop
$\gamma_o$ is $\Psi_{\gamma_o}(\gamma)$, the characteristic function of the
loop $\gamma_o$, that we introduced in section 3.3. This is the eigenstate
of the electric field operators (appropriate to the specific loop
representation under consideration) $\hat{E}(f)$ with eigenvalues $i\hbar
\oint_\gamma f_a dS^a$. These states represent the simplest excitations of
the electric field. Why are the excitations associated with closed loops
$\gamma_o$ rather than points? In the quantum theory of scalar fields, for
example, the simplest excitations {\it are} localized at points. Why is the
same not true of the electric field? The reason is that the electric field
is constrained to satisfy the Gauss law and divergence-free vector fields
cannot be supported on points. The simplest geometrical structure on which
they can have support are precisely closed loops! Thus, Faraday's vision of
the electromagnetic field in terms of lines of force is realized in two
different ways in the in the loop description of quantum electro-magnetism.
\bigskip
\goodbreak

{\bf Acknowledgements}: We have benefited from discussions with a number of
colleagues. Among them, we would especially like to thank Chris Isham and
Lee Smolin. This work was supported in part by the NSF grants
INT88-15209, PHY90-12099 and PHY90-16733; by the research funds provided
by Syracuse University; and, by a SERC Visiting Fellowship (to AA).

%\newpage

\end{document}